\documentclass[ reprint,superscriptaddress, amsmath,amssymb, aps, longbibliography]{revtex4-1}

\usepackage[colorlinks=true,linkcolor=blue,urlcolor=blue,citecolor=blue]{hyperref}
\usepackage{overpic}
\usepackage{upgreek}
\usepackage{physics}
\usepackage{graphicx}
\usepackage{dcolumn}
\usepackage{bm}
\usepackage{outlines}
\usepackage[dvipsnames]{xcolor}
\usepackage{gensymb}
\usepackage{subcaption}
\usepackage{caption}
\usepackage{calc}
\usepackage{tikz}  
\usepackage{comment}
\usepackage{pdfpages}
\makeatletter
\AtBeginDocument{\let\LS@rot\@undefined}
\makeatother

\newcommand{\mitll}{MIT Lincoln Laboratory, Lexington, MA 02421, USA}

\usepackage{ulem}

\begin{document}

\title{Flux-trapping characterization for superconducting electronics using a cryogenic widefield N-$V$ diamond microscope}
\author{Rohan T.~Kapur}
\affiliation{\mitll}
\author{Pauli Kehayias}
\affiliation{\mitll}
\author{Sergey K.~Tolpygo}
\affiliation{\mitll}
\author{Adam A.~Libson}
\affiliation{\mitll}
\author{George Haldeman}
\affiliation{\mitll}
\author{Collin N. Muniz}
\affiliation{\mitll}
\author{Alex Wynn}
\affiliation{\mitll}
\author{Nathaniel J.~O'Connor}
\affiliation{\mitll}
\author{Neel A.~Parmar}
\affiliation{\mitll}
\author{Ryan Johnson}
\affiliation{\mitll}
\author{Andrew C.~Maccabe}
\altaffiliation{Current Affiliation: Quantum Science and Engineering Program, Harvard University, Cambridge, MA 02138, USA}
\affiliation{\mitll}
\author{John Cummings}
\affiliation{\mitll}
\author{Justin L.~Mallek}
\affiliation{\mitll}
\author{Danielle A.~Braje}
\affiliation{\mitll}
\author{Jennifer M.~Schloss}
\affiliation{\mitll}


\begin{abstract}
Magnetic flux trapping is a significant hurdle limiting the reliability and scalability of superconducting electronics, yet tools for imaging flux vortices remain slow or insensitive. We present a cryogenic widefield NV-diamond magnetic microscope capable of rapid, micrometer-scale imaging of flux trapping in superconducting devices. Using this technique, we measure vortex expulsion fields in Nb thin films and patterned strips, revealing a crossover in expulsion behavior between $10$ and $20~\upmu$m strip widths. The observed scaling agrees with theoretical models and suggests the influence of film defects on vortex expulsion dynamics. This instrument enables high-throughput magnetic characterization of superconducting materials and circuits, providing new insight for flux mitigation strategies in scalable superconducting electronics.
\end{abstract}
\maketitle

\section{Introduction}
Superconducting computing is a compelling alternative to complementary metal-oxide semiconductor (CMOS) technologies, offering faster clock speeds and dramatically improved power efficiency~\cite{SCEreviewVanDuzer, SCEreviewSpringer, SCEreviewIEEE2024}. Superconductors enable low-loss, dispersion-free data transmission at frequencies up to $\sim$500 GHz. For classical digital logic, Josephson-junction-based superconducting electronics (SCE) achieve the lowest energy per bit, approaching the Landauer limit. Even after accounting for cryogenic cooling, SCE systems promise $\sim$100$\times$ lower energy consumption than CMOS, and clock speeds exceeding 100 GHz\textemdash{}far beyond the few-GHz ceiling for conventional electronics~\cite{rsfq770GHz}.

Despite these advantages, achieving very-large-scale integration (VLSI) with SCE requires overcoming the persistent challenge of magnetic flux trapping~\cite{moatsIEEE, evalOfFluxTrapping}. SCE devices rely on type-II superconductors, which trap quantized magnetic flux (vortices) when cooled below their critical temperature $T_c$ in a magnetic field. Vortices near sensitive components, such as Josephson junctions, can disrupt circuit operation. Mitigation strategies\textemdash{}including minimizing the magnetic field during cooldown, patterning ``moats" (etched holes and slots in superconducting films) to attract flux away from critical components, applying thermal gradients to reposition vortices, and vortex ratcheting with AC signals\textemdash{}have shown promise~\cite{moatsIEEE, thermalGrad, ratchetEffect}. However, the absence of a fast, reliable characterization technique for flux trapping has hindered efforts to optimize these methods, leaving the magnetic flux trapping problem largely unresolved.

Pinpointing vortex locations and understanding their effects on SCE circuit components is crucial for developing effective mitigation. Single-flux-quantum-based shift registers have been used to electrically detect trapped flux via error syndromes~\cite{shregNew, shregOld}. While these diagnostics can identify affected logic cells, they cannot localize individual vortices or resolve sub-cell-scale interactions, and their operation can also become compromised by flux trapping.

Magnetic field imaging (MFI) offers a complementary, spatially resolved approach to flux characterization. By directly visualizing vortex locations, MFI provides insight into where vortices typically form and how their locations correlate with device failure modes. Several MFI techniques have successfully imaged vortices in superconducting films and circuits~\cite{kirtley2010review, initial_sc_strip_paper, moiDOEref, maletinksyVortices, aniaVorticies}. While these methods have proven useful, their slow acquisition times (as long as a day to measure one device), poor signal-to-noise ratios (SNRs), and/or limited field-of-view (FOV) sizes have prevented their widespread adoption for practical SCE diagnostics. The design space of SCE devices and trapped-flux mitigation techniques is extensive, and effective troubleshooting requires fast, reliable, high-throughput measurements over many cooldown cycles. A key performance metric for MFI-based characterization of SCE is the area measurement rate, $\dot{A}$, defined as the areal speed at which superconducting samples containing vortices can be imaged with an acceptable SNR ($\geq 5$)~\cite{suppl}.

\begin{figure*}[htbp]
  \centering
  \captionsetup{font=small}

  \newsavebox{\gridbox}
  \newlength{\gridheight}

  \sbox{\gridbox}{%
    \begin{minipage}{0.68\textwidth}
      \centering

      \begin{overpic}[width=0.49\linewidth, trim=0pt 0pt 0pt 0pt, clip]{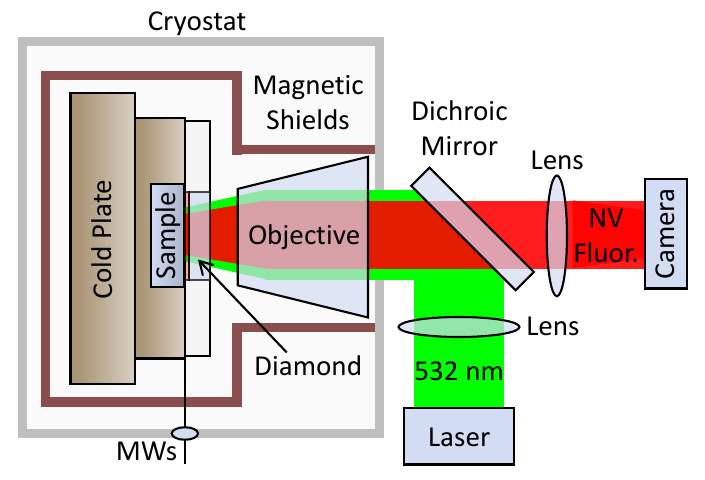}
        \put(0,66){\footnotesize\textbf{(a)}}
      \end{overpic}\hfill
      \begin{overpic}[width=0.49\linewidth, trim=0pt -15pt 5pt 0pt, clip]{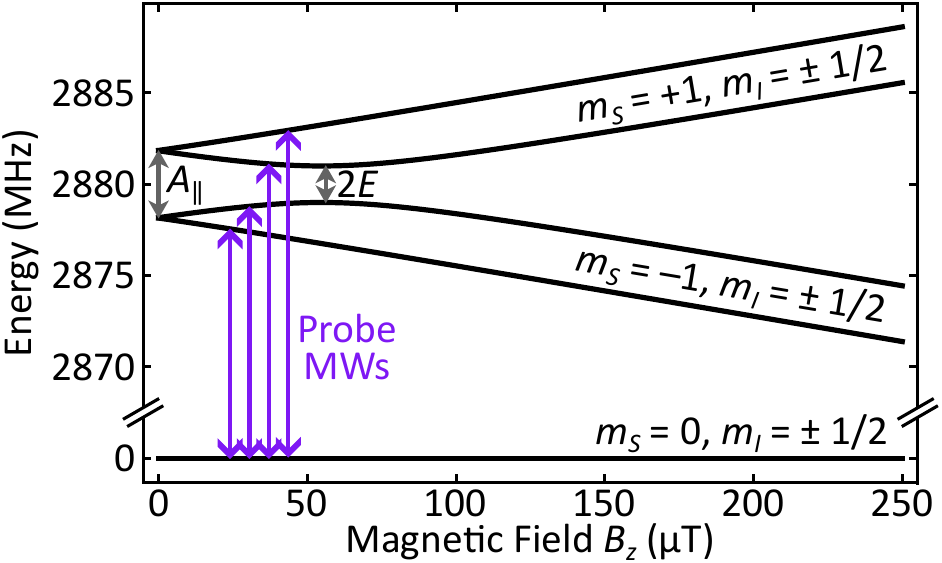}
        \put(0,68.4){\footnotesize\textbf{(b)}}
      \end{overpic}

      \vspace{2mm}

      \begin{overpic}[width=0.49\linewidth, trim=0pt 0pt 0pt 0pt, clip]{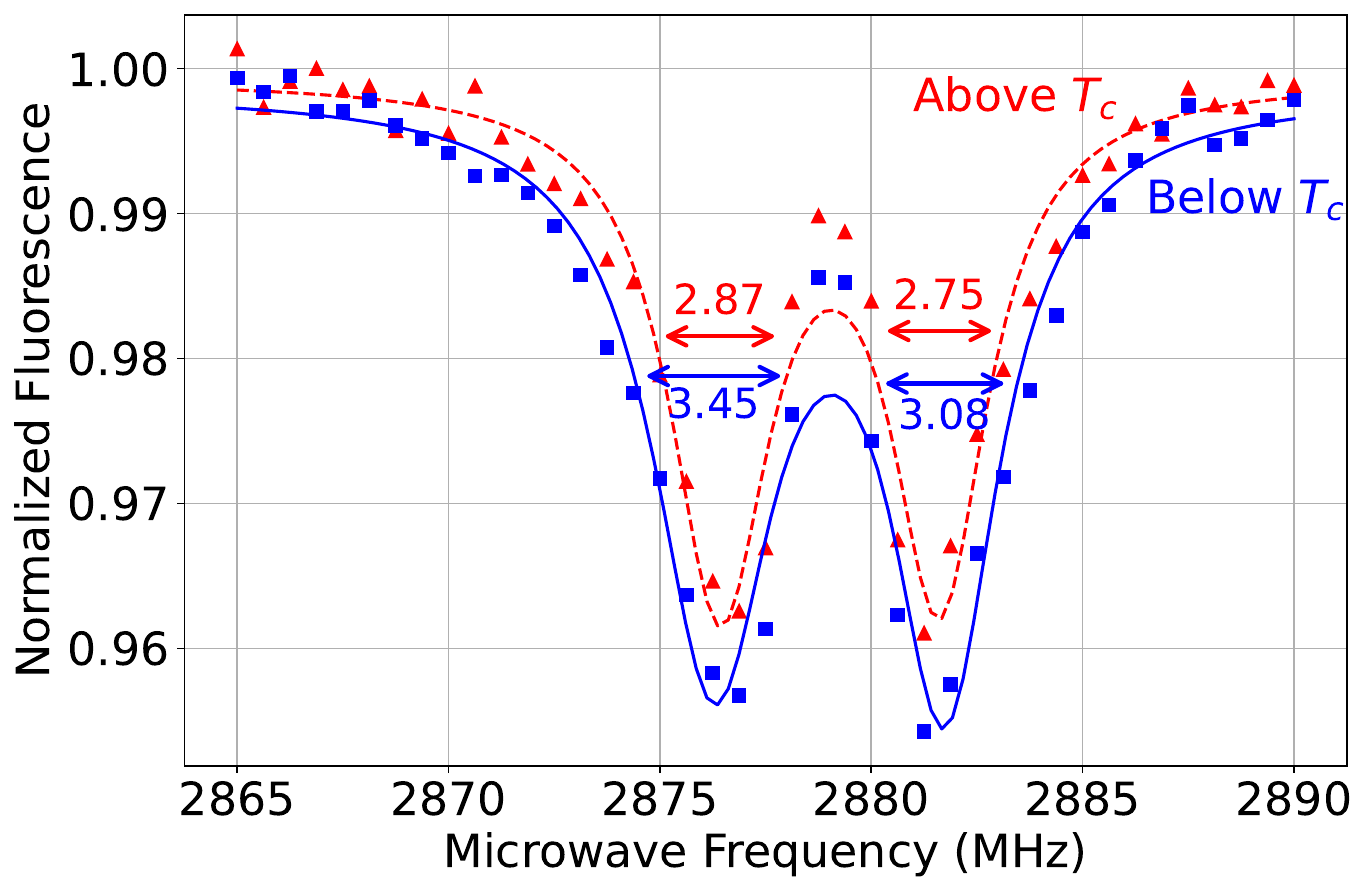}
        \put(0,64.2){\footnotesize\textbf{(c)}}
      \end{overpic}\hfill
      \begin{overpic}[width=0.49\linewidth, trim=0pt 0pt 5pt 0pt, clip]{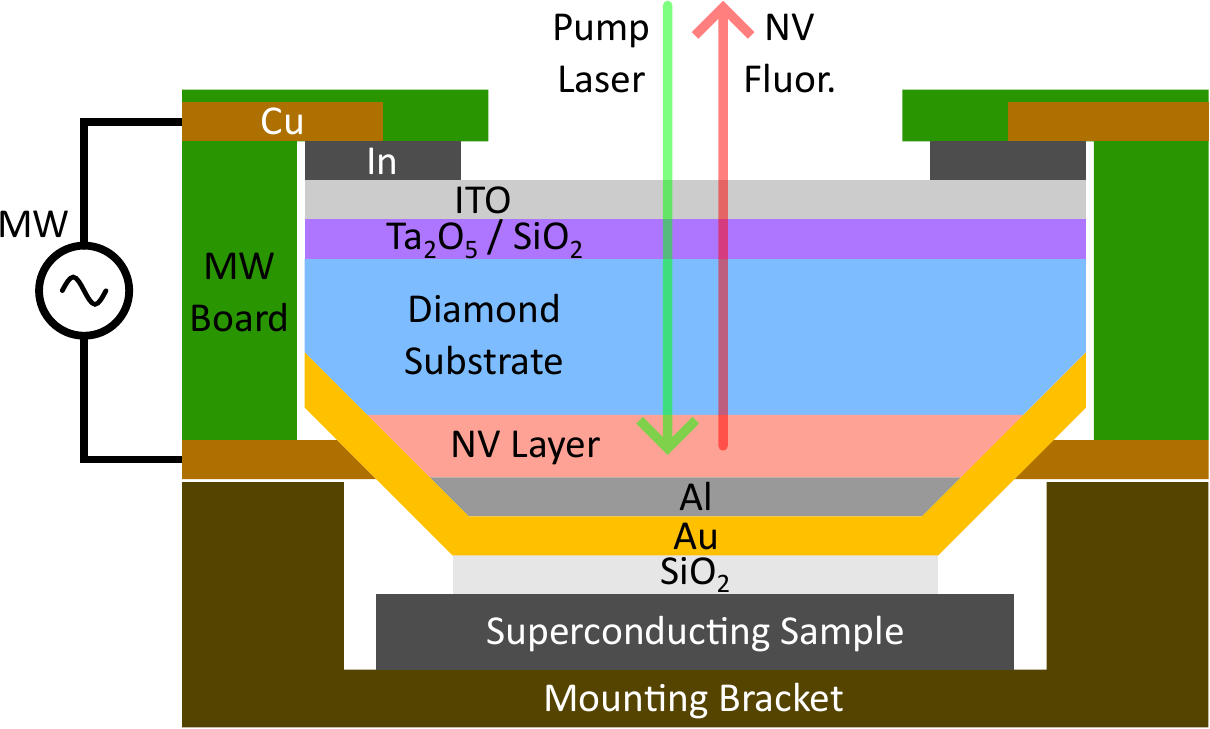}
        \put(0,66){\footnotesize\textbf{(d)}}
      \end{overpic}
    \end{minipage}%
  }

  \setlength{\gridheight}{\ht\gridbox}
  \addtolength{\gridheight}{\dp\gridbox}

  \usebox{\gridbox}\hfill
  \begin{minipage}{0.29\textwidth}
    \centering
    \begin{overpic}[width=\linewidth, height=\gridheight, keepaspectratio,
      trim=0pt 0pt 0pt 0pt, clip]{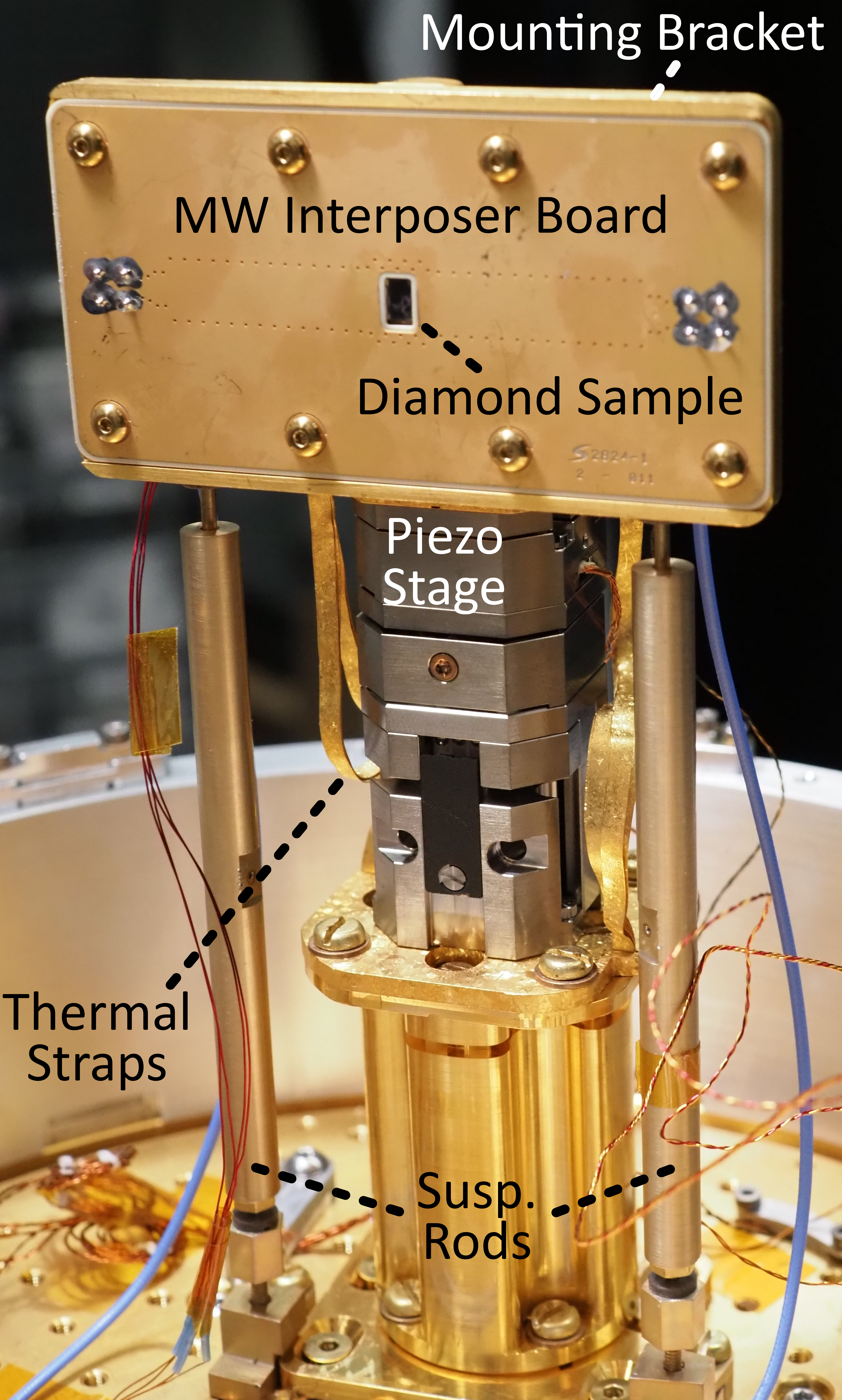}
      \put(-6,100.2){\footnotesize\textbf{(e)}}
    \end{overpic}
  \end{minipage}

  \caption{(a) Schematic of the NV cryo-microscope. (b) NV ground-state energy levels as a function of on-axis magnetic field $B_z$. (c) Example single-pixel ODMR spectra measured at a vortex location. The vortex magnetic field below $T_c$ results in an observed increase in the ODMR linewidth $\Gamma$. The increased fluorescence contrast observed can be attributed to temperature-dependent effects~\cite{low_temp_NV_effects}. (d) Cross-section view showing the diamond optical coatings and overall diamond/sample stackup. (e) A photograph of the sensor head showing the various hardware components used for sample positioning, MW delivery, and thermal management.}
  \label{fig:apparatus_figure}
\end{figure*}

In this work, we present a widefield cryogenic nitrogen-vacancy (NV) diamond magnetic microscope designed for rapid, high-resolution imaging of flux vortices in superconducting devices~\cite{victorNVscReview, heziVortices, nishimuraYBCOvortices}. To achieve this, our system images a 2.5~mm $\times$ 4.5~mm chip area with micrometer-scale resolution, operates from room temperature to $\sim$4 K, and functions in background fields from below $10$~nT to $1$~mT, all while isolating samples from the laser light and microwave (MW) control fields. These features enable rapid assessment of flux trapping across multiple cooldowns, allowing for quantitative evaluation of flux mitigation strategies with a typical image taking 4 minutes for a 576 $\upmu$m $\times$ 360 $\upmu$m FOV ($\dot{A} = 860~\upmu$m$^2$/s), which surpasses previous approaches. We demonstrate localization of individual vortices, identification of persistent pinning sites in unpatterned Nb films, and measurements of vortex expulsion fields in patterned superconducting strips and films with moats. The expulsion field is the maximum residual out-of-plane magnetic field under which a structure can be cooled through $T_c$ without trapping flux\textemdash{}a critical parameter for scalable SCE and type-II superconductor physics. This instrument lays the foundation for systematic studies of flux dynamics in increasingly complex film geometries, multilayer stacks, and ultimately active superconducting circuits.

\section{Instrumentation and Measurement Protocol}
The experimental setup consists of a room-temperature widefield optical microscope coupled to a 4 K closed-cycle cryostat that houses the NV-diamond sensor and superconducting sample (Fig.~\ref{fig:apparatus_figure}a). Magnetic field images are acquired using continuous-wave optically detected magnetic resonance (CW-ODMR) spectroscopy from a layer of $^{15}$NV centers near the diamond surface \cite{QDM1ggg, edlynQDMreview}. The NV layer is continuously illuminated with 50\textendash{}90 mW of 532 nm laser light delivered through a 20$\times$ microscope objective lens maintained at room temperature inside the cryostat.  The laser light continuously optically pumps the NVs into the strong-fluorescence $m_S = 0$ ground-state sublevel (Fig.~\ref{fig:apparatus_figure}b).  NV fluorescence is collected through the same objective and imaged onto a machine vision CMOS camera. The single-shot FOV, with dimensions 360~$\upmu$m~$\times$~576~$\upmu$m, can be tiled to image a 2.5~mm~$\times$~4.5~mm sample area.   A probe MW field drives transitions to the weaker-fluorescence $m_S = \pm 1$ sublevels.  Local magnetic fields along the NV orientation axes shift the NV spin transition frequencies, altering the detected ODMR spectra at each pixel.  The diamonds used in the apparatus are engineered to have a 1~$\upmu$m thick NV layer and $\sim$500 nm global flatness, enabling imaging of micrometer-scale magnetic features (see Appendix \ref{diamondSpecsAppendix}). Additional details on the MW interposer board design, NV Hamiltonian near zero magnetic field, and area measurement rate assessment are provided in the Supplemental Material~\cite{suppl}.

To image a superconducting sample, we sweep the MW frequency across the ground-state transitions, capture fluorescence images at each frequency, and fit the resulting ODMR spectra at each pixel using a dual-Lorentzian function (Fig.~\ref{fig:apparatus_figure}c). Near zero magnetic field, the transition frequencies are given by $D \pm \sqrt{(A_{\parallel}/2 \pm \gamma B_z)^2 + E^2}$, where $D \approx 2880$ MHz and $E$ are zero-field splitting parameters, $A_{\parallel} \approx 3.065$ MHz is a hyperfine coupling constant for $^{15}$N, $\gamma \approx 28$ GHz/T is the NV gyromagnetic ratio, and $B_z$ is the magnetic field along the NV axis~\cite{QDM1ggg, n15vSplittingsNearZeroField, n14n15HFvsTemperature, suppl}. Magnetic field signals manifest as additional line broadening and splitting in the ODMR spectrum. The precise changes in broadening and splitting in response to changes in magnetic field depend on the NV ensemble properties, making the choice of whether to use broadening or splitting to measure magnetic field diamond-specific. For the particular diamonds used in this work, we used lineshape broadening to extract the magnetic field information. Each magnetic image typically requires a few minutes to acquire and spans a 576~$\upmu$m~$\times$~360~$\upmu$m FOV.

For each diamond, the NV side was coated with a reflective Al layer to prevent light from reaching the superconducting sample, an Au layer to form an electrical contact for microwave delivery, and a SiO$_2$ insulating layer to prevent electrical shorts between the diamond and the sample (Fig.~\ref{fig:apparatus_figure}(d)). The non-NV side was coated with an indium tin oxide (ITO) layer to form a transparent electrical connection, and an anti-reflective Ta$_2$O$_5$/SiO$_2$ coating. The NV surface was chamfered along two edges for electrical interfacing with the MW interposer board. Indium foil was used to ensure electrical contact between the diamond and the board.

The diamond and superconducting sample are mounted to a sensor head, which consists of the MW interposer board affixed to a copper right-angle mounting bracket (Fig.~\ref{fig:apparatus_figure}e). The bracket includes a sample holder (compatible with chips up to 5~mm~$\times$~5~mm), a magnetic field compensation coil capable of generating out-of-plane fields up to 1~mT, and a thin-film temperature sensor. The sensor head compresses the NV layer against the sample to minimize standoff distance and provide thermal contact with the cold plate. To maintain mechanical and thermal stability, the sensor head includes thermal straps and suspension rods. It is mounted on a non-magnetic three-axis piezoelectric stage for focus and lateral translation of the FOV, and enclosed in magnetic shielding with remnant field $<100$~nT. The full assembly is housed in a closed-cycle optical cryostat operating from room temperature to 4~K (Fig.~\ref{fig:apparatus_figure}a).

Accurate flux trapping diagnostics with a widefield NV-diamond cryo-microscope require a near-zero background magnetic field and effective shielding of the superconducting sample from the laser and microwave (MW) fields used to interrogate the NV layer. A high-reflectivity coating on the sample-facing side of the diamond protects the superconductor from laser light, while conductive coatings on both sides of the diamond confine the MW fields~\cite{suppl}. Unlike standard NV magnetometry setups that use unconfined MW delivery (e.g., shorted coaxial loops), we employ an integrated interposer board in which the diamond acts as a capacitive element in a microstrip circuit~\cite{samQFMimaging}. This design ensures uniform MW excitation across the FOV while suppressing stray MW fields at the sample to below 100 nT. The sample is mounted within nested magnetic shields along with an embedded coil for applying controlled out-of-plane magnetic fields.

Measurements are performed above and below the superconducting sample critical temperature ($T_c \approx 9.2$ K for Nb), typically at 10 K and 5 K, respectively, with the applied magnetic field fixed during cooldown and measurements. At each pixel, we extract the ODMR linewidth above ($\Gamma_{T>T_c}$) and below $T_c$ ($\Gamma_{T<T_c}$), and compute their difference $\Delta\Gamma = \Gamma_{T<T_c} - \Gamma_{T>T_c}$. This differential imaging approach isolates superconductor-related signals and suppresses artifact sources such as diamond strain features. To convert $\Delta\Gamma$ to a magnetic field, we first calibrate the dependence of $\Gamma$ on applied field above $T_c$, then interpolate the observed $\Delta\Gamma$ values using the resulting calibration curve.

\begin{figure*}[htbp]
  \centering
  \captionsetup{font=small} 

  \begin{minipage}{0.33\textwidth}
    \centering
    \begin{overpic}[trim=50pt 50pt 0pt 45pt, clip, width=\textwidth]{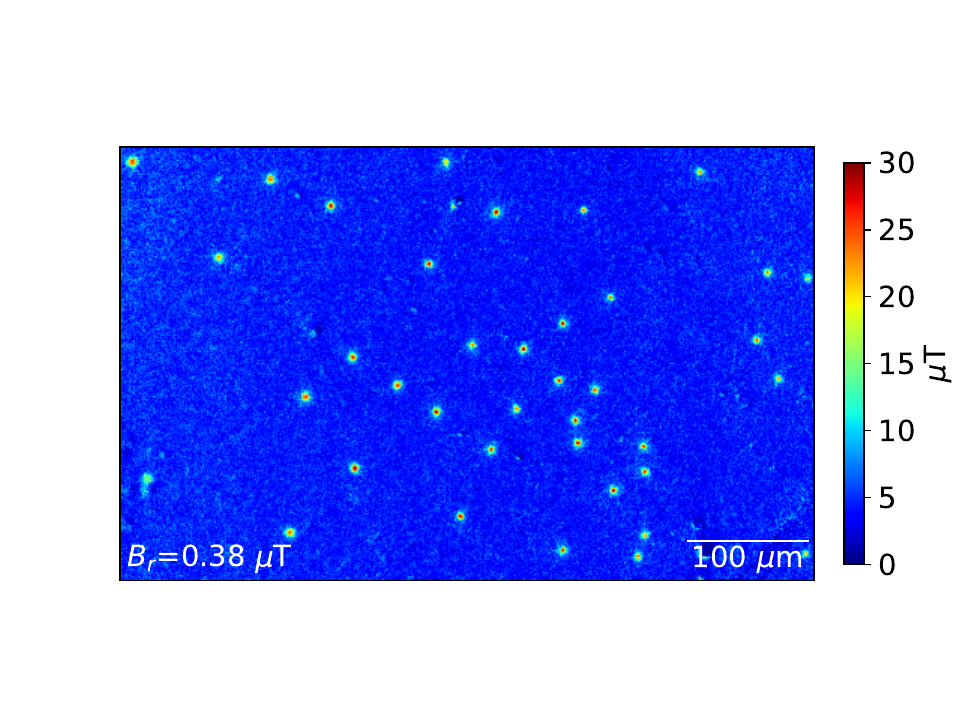}
      \put(-2,58){\footnotesize\textbf{(a)}}
    \end{overpic}
    \label{fig:low_field_bare_film}
  \end{minipage}%
  \begin{minipage}{0.33\textwidth}
    \centering
    \begin{overpic}[trim=50pt 45pt 0pt 45pt, clip, width=\textwidth]{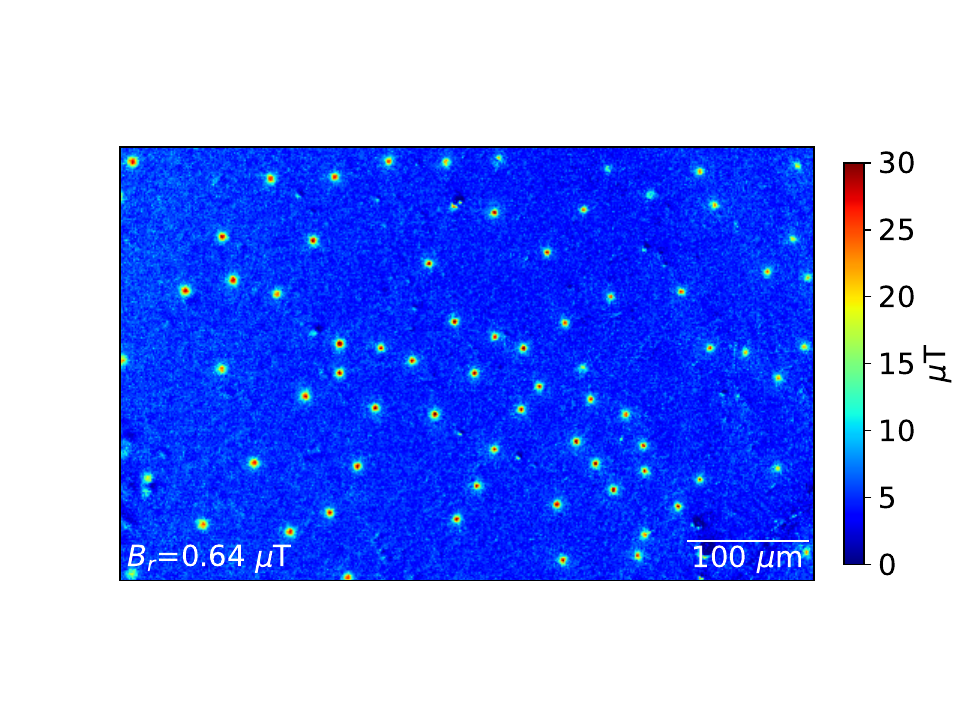}
      \put(-2,58){\footnotesize\textbf{(b)}}
    \end{overpic}
    \label{fig:temp_cycle_high_field_bare_film}
  \end{minipage}%
  \begin{minipage}{0.33\textwidth}
    \centering
    \begin{overpic}[width=\columnwidth, trim=70pt 70pt 0pt 10pt, clip]{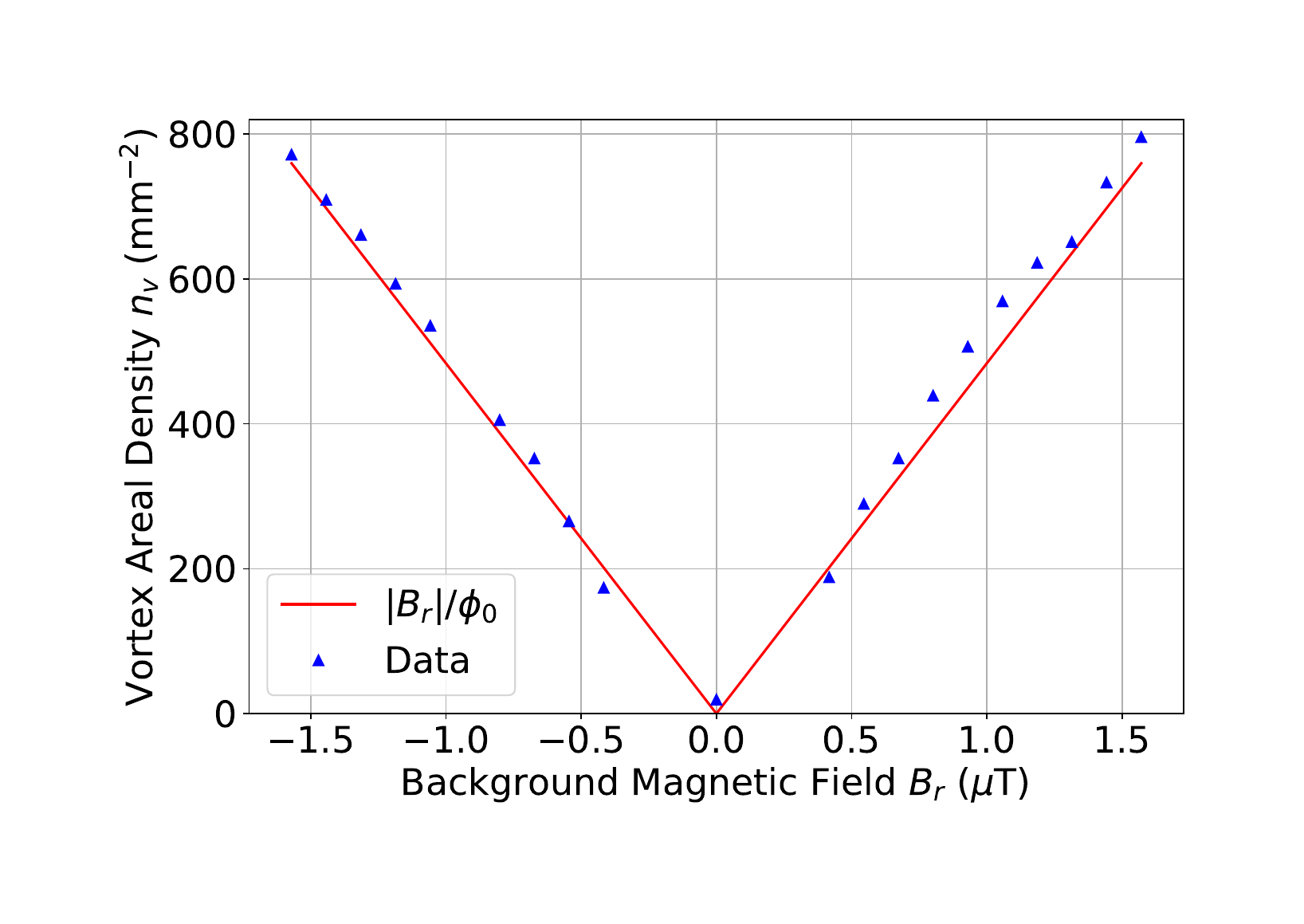}
      \put(-2, 59){\footnotesize\textbf{(c)}}
    \end{overpic}
    \label{fig:vortex_ODMR}
  \end{minipage}

  \vspace{-1em}

  \begin{minipage}{0.33\textwidth}
    \centering
    \begin{overpic}[trim=50pt 50pt 0pt 50pt, clip, width=\textwidth]{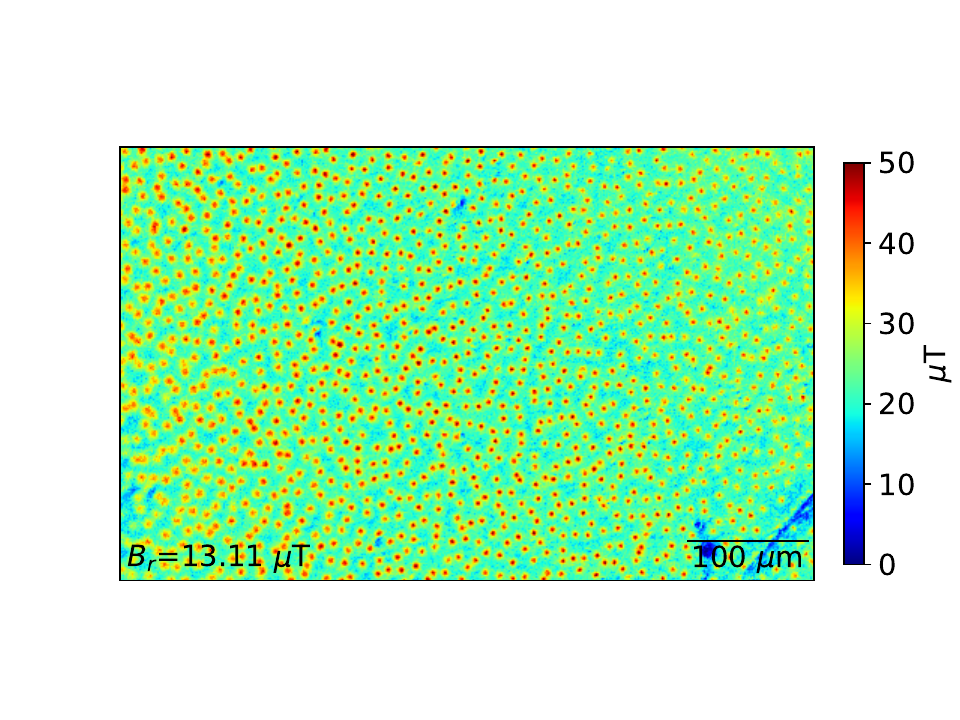}
      \put(0,58){\footnotesize\textbf{(d)}}
    \end{overpic}
    \label{fig:2mABareFilm}
  \end{minipage}%
  \begin{minipage}{0.33\textwidth}
    \centering
    \begin{overpic}[trim=50pt 50pt 0pt 50pt, clip, width=\textwidth]{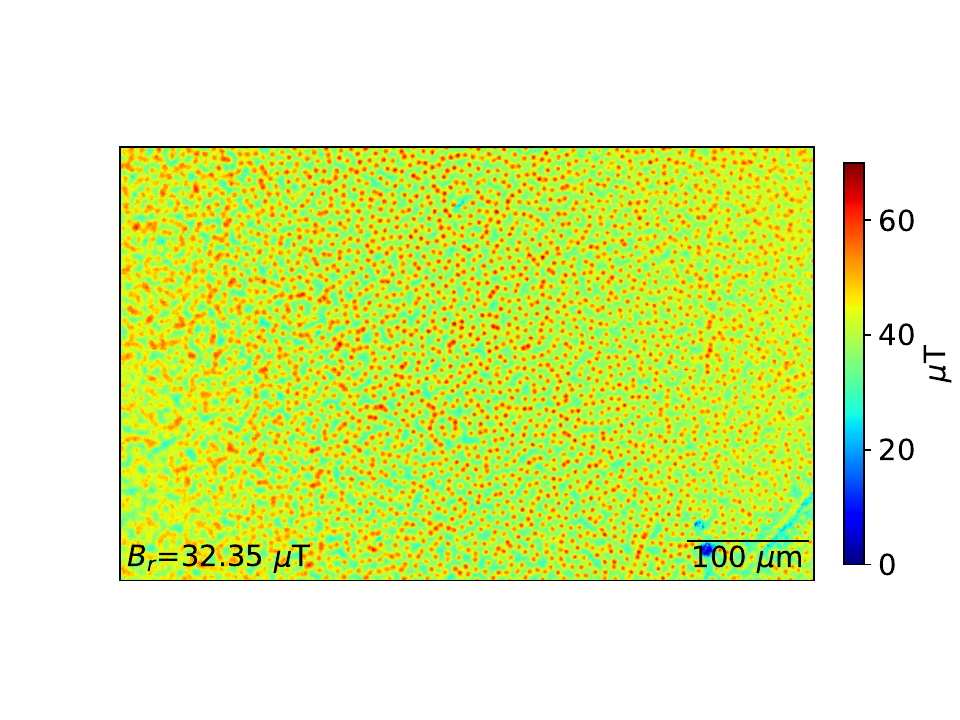}
      \put(-2,58){\footnotesize\textbf{(e)}}
    \end{overpic}
    \label{fig:5mABareFilm}
  \end{minipage}%
  \begin{minipage}{0.33\textwidth}
    \centering
    \begin{overpic}[trim=50pt 50pt 0pt 50pt, clip, width=\columnwidth]{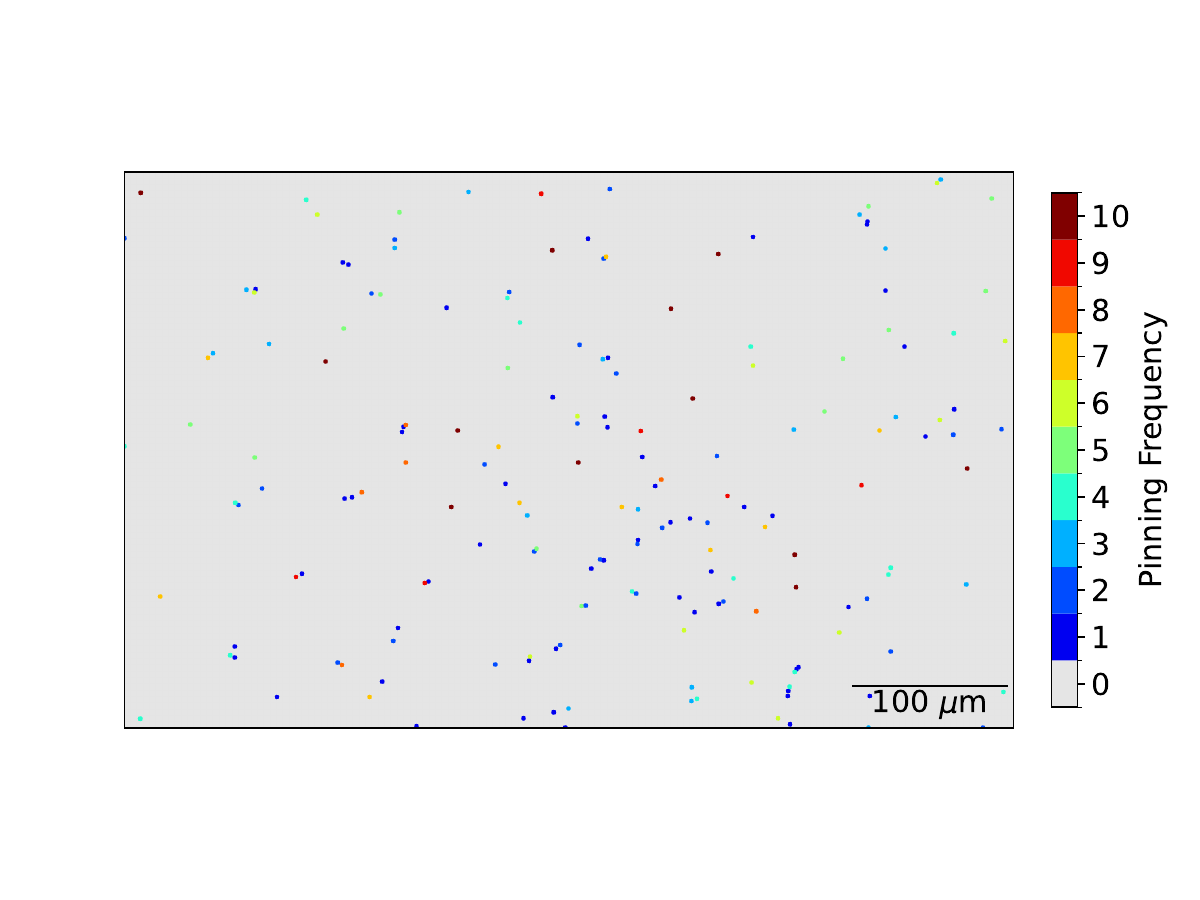}
      \put(0,60){\footnotesize\textbf{(f)}}
    \end{overpic}
    \label{fig:non_vortex_ODMR}
  \end{minipage}

  \vspace{-2em}
    
  \caption{(a)-(b) Magnetic field images showing flux vortices in a bare superconducting Nb film in background fields $B_r=$ 0.38~$\upmu$T and 0.64~$\upmu$T. (c) Measured vortex areal density over a range of $B_r$. (d)-(e) Magnetic field images of the film at $B_r=$ 13.11~$\upmu$T and 32.35~$\upmu$T. (f) Map of vortex pinning sites observed across ten temperature cycles at $B_r=$ 0.64~$\upmu$T, where dot color indicates the frequency with which vortices appeared. An example image from this dataset is shown in (b).}
  \label{fig:nb_bare_film_images}  
\end{figure*}

\section{Results}
\subsection{Vortices in an Unpatterned Nb Film}
To validate the system performance, we imaged vortices in unpatterned Nb test structures fabricated using Nb deposition and etching parameters identical to the MIT Lincoln Laboratory SFQ5ee process~\cite{SFQ5ee, suppl}. A 2.5 mm~$\times$~2.5 mm~$\times$~200 nm Nb film on oxidized Si was cooled through $T_c$ to $\sim$5 K in a range of background fields $B_r$ from 10 nT to 45~$\upmu$T. For $|B_r| \leq 1.6~\upmu$T (e.g., in Fig.~\ref{fig:nb_bare_film_images}a,b), we tracked the number of observed vortices $N$ in the FOV area $A$ (360~$\upmu$m~$\times$~576~$\upmu$m) and computed the vortex areal density $n_v = N/A$. As shown in Fig.~\ref{fig:nb_bare_film_images}c, $n_v$ exhibits the expected linear relation $n_v = |B_r|/\Phi_0$ (where $\Phi_0 = h/2e \approx 2.07 \times 10^{-15}~\mathrm{T \cdot m^2}$ is the magnetic flux quantum) measured down to a single vortex when nulling the remnant magnetic field to $B_{r}\approx10$ nT.

As seen at larger values of $B_r$ (Fig.~\ref{fig:nb_bare_film_images}d,e), the observed vortex distributions do not form ordered lattices, as would be expected in an ideal type-II superconductor\textemdash{}likely due to strong pinning centers in the film. To quantify flux pinning behavior, we imaged the same FOV across ten cooldown cycles at $B_r = 0.64$~$\upmu$T. Vortex locations were identified using a blob-counting algorithm~\cite{van2014scikit}, and sites that repeatedly hosted vortices across cooldowns were cataloged. Figure~\ref{fig:nb_bare_film_images}f maps these recurrent pinning sites, color-coded by vortex occurrence frequency. Across the ten cooldowns, we identified 180 unique pinning sites, with each image containing an average of 67 vortices (Video~\ref{video1}).

\begin{video}[h]
    \centering
    \includegraphics[
        width=1.03\columnwidth,
        trim=2.3cm 2.3cm 0cm 2.3cm,  
        clip
    ]{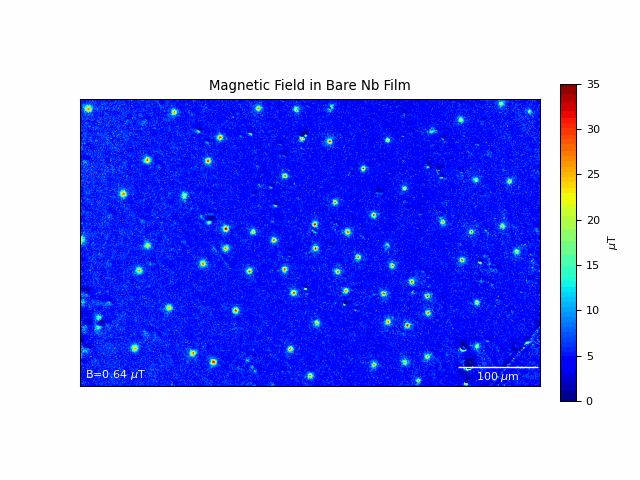}
    \caption{\label{video1} A video showing flux trapping in a Nb film across ten cooldowns, with $B_r=$ 0.64~$\upmu$T.}
\end{video}

\begin{figure}[htbp]
    \centering
    \vspace{0.5em}
    \begin{overpic}[width=0.9\linewidth, trim=0pt 0pt 0pt 0pt, clip]{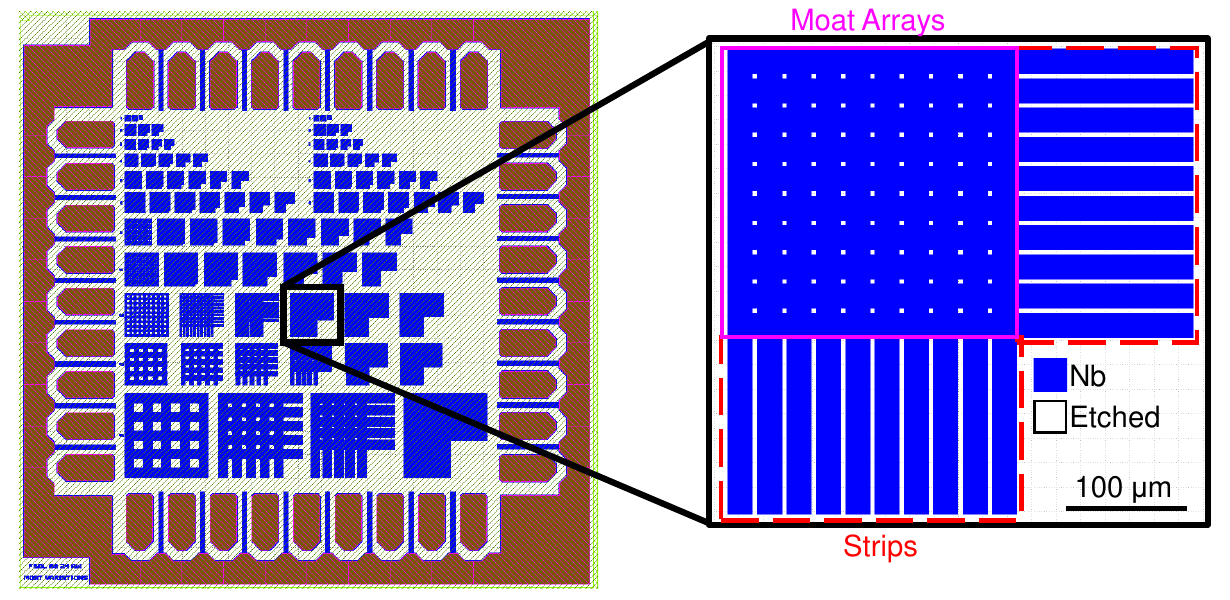}
        \put(-7, 45){\footnotesize\textbf{(a)}}
    \end{overpic}
\vspace{-.5em}
    \begin{overpic}[width=0.9\linewidth, trim=50pt 30pt 5pt 10pt, clip]{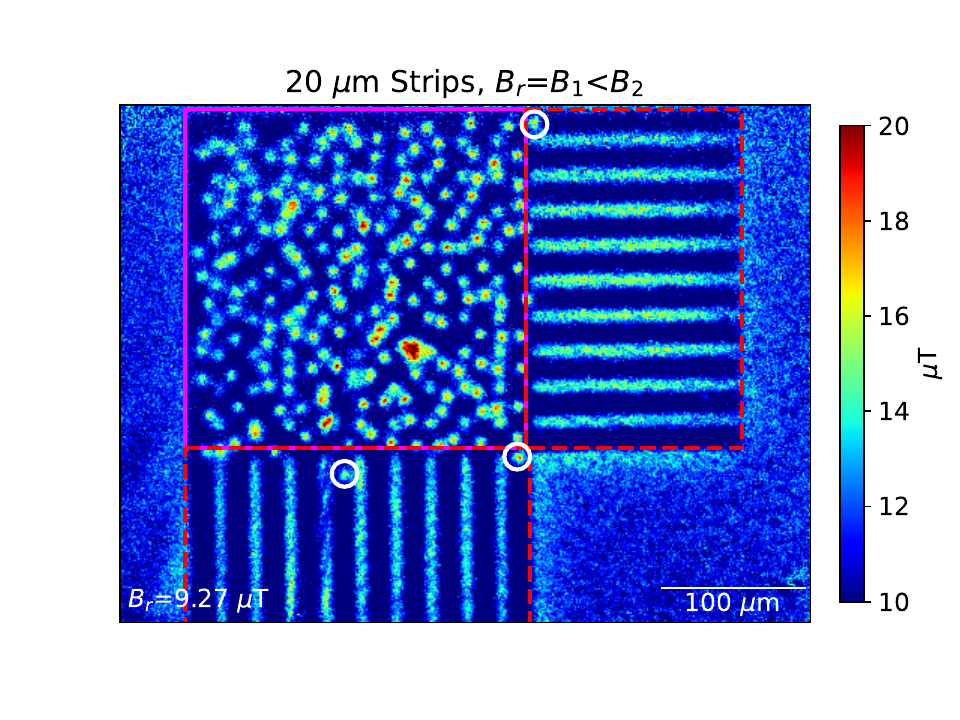}
        \put(-7, 67){\footnotesize\textbf{(b)}}
    \end{overpic}
\vspace{-.5em}
    \begin{overpic}[width=0.9\linewidth, trim=50pt 30pt 5pt 10pt, clip]{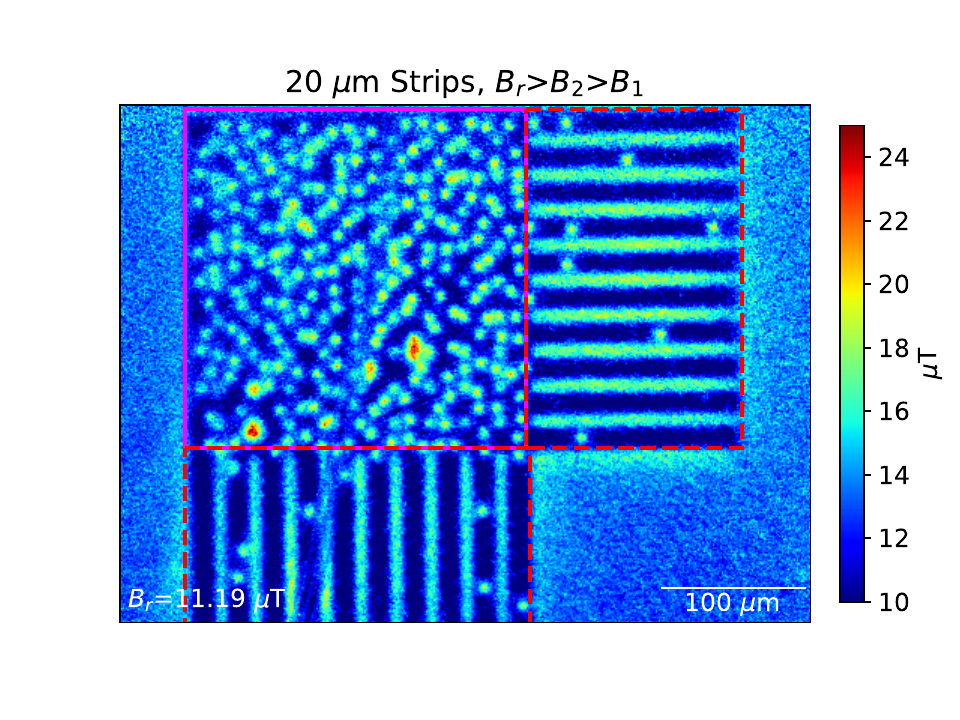}
        \put(-7, 67){\footnotesize\textbf{(c)}}
    \end{overpic}

    \caption{(a) Test chip layout with 20 $\upmu$m strip test structure highlighted. The blue regions are Nb, while in the white regions the superconductor has been etched away. The outlined red-dashed region is the strip region, where all strip expulsion field measurements were performed, while the outlined solid-magenta region is the moat-array region. (b) 20 $\upmu$m strips for $B_r=B_{1}<B_2$, where the first vortices in strips are observed (circled in white). (c) 20 $\upmu$m strips for $B_r>B_{2}>B_1$, with many vortices observed in the strips.}
    \label{fig:critical_field_measurement}
\end{figure}

\subsection{Vortices in Patterned Nb Films} 
Understanding the vortex expulsion field $B_{\mathrm{exp}}$ in patterned superconducting films is essential for designing flux-robust SCE. The expulsion field is the maximum magnetic field below which a superconducting structure can be cooled through $T_c$ without trapping vortices, a parameter that places practical constraints on circuit layout and shielding requirements. To systematically study vortex expulsion behavior in geometries relevant to superconducting electronics, we fabricated a 5~mm $\times$ 5~mm test chip containing patterned Nb structures (Fig.~\ref{fig:critical_field_measurement}a). These structures are composed of a 200 nm-thick Nb layer on oxidized Si, and include arrays of perpendicular strips with variable widths $W$ (1\textendash{}80~$\upmu$m) and spacings (1\textendash{}80~$\upmu$m). The design allows $B_{\mathrm{exp}}$ measurements spanning roughly three orders of magnitude using a single chip with consistent deposition and etching conditions. Each array contains two types of regions: (a) isolated strips, which are straightforward to interpret and enable comparison to prior work~\cite{initial_sc_strip_paper} (outlined in red in Fig.~\ref{fig:critical_field_measurement}a); and (b) a square film with an array of etched square moats (antidots) of side length $a$ and spacing $s$, forming a so-called ``Swiss cheese" pattern (outlined in magenta in Fig.~\ref{fig:critical_field_measurement}a).

Here we focus primarily on the strips due to their geometric simplicity and direct comparability to prior studies~\cite{initial_sc_strip_paper}. We measured $B_{\mathrm{exp}}$ values for strips with widths ranging from 4 to 80~$\upmu$m. The first expulsion field, $B_1$, is defined as the applied field $B_r$ above which vortices first appear in the strip, likely due to trapping at pinning sites. The second expulsion field, $B_2$, marks the onset of a linear increase in vortex areal density $n_v$ with a slope of $\Phi_0^{-1}$, consistent with the behavior of an unpatterned film. We extract $B_2$ by extrapolating this linear regime back to $n_v = 0$ (Fig.~\ref{fig:combined_critical_field_figures}a)~\cite{suppl}.

To determine $B_1$ and $B_2$, we measured $n_v$ as a function of $B_r$ for each strip width~\cite{suppl}. Figure~\ref{fig:combined_critical_field_figures}a shows representative $n_v$ curves for 10, 20, and 40~$\upmu$m-wide strips. The field dependence differs qualitatively between wide and narrow strips. For wider strips (40 and 80~\(\upmu\)m), we observe two distinct regimes: complete flux expulsion below $B_1$, followed by a sudden linear increase in $n_v$ with a slope of approximately $\Phi_0^{-1}$ for $ B_r > B_1 $. For these strips, $B_2$, defined by extrapolating this linear regime to $n_v = 0$, is approximately equal to $B_1$. In narrower strips (4-20~$\upmu$m), the field dependence splits into three regimes: (i) no vortices for $ B_r < B_1 $; (ii) a slow, approximately linear increase in $n_v$ with slope less than $\Phi_0^{-1}$; and (iii) a steeper, linear increase in $n_v$ with slope ${\sim}\Phi_0^{-1}$. As before, $B_2$ is extracted by extrapolating the steep linear regime to $n_v = 0$. In regime (ii), vortices tend to recur at the same positions, consistent with strong pinning sites \cite{suppl}. Figure~\ref{fig:combined_critical_field_figures}b shows the measured $B_1$ and $B_2$ values as a function of strip width, alongside predictions from several theoretical models examined in the Discussion.

In the course of imaging the strips, we also observed flux trapping in the square moat array regions of the measured chip, characterized by side length $a$, spacing $s$, and pitch $a + s$, with an example region outlined in magenta in Fig.~\ref{fig:critical_field_measurement}a-c. We observed that vortices between moats with spacing $s$ systematically appeared at lower magnetic fields than in the strips with width $W=s$ for all moat sizes $a$. The exact value of expulsion field for the film with moats depends on both $a$ and $s$~\cite{suppl}, and a detailed analysis of flux expulsion in these moat geometries is reserved for a separate study \cite{nvMoatsPaper}.

\section{Discussion}
\begin{figure}[h]
    \centering
    \vspace{0.5em}
    \begin{overpic}[width=\linewidth]{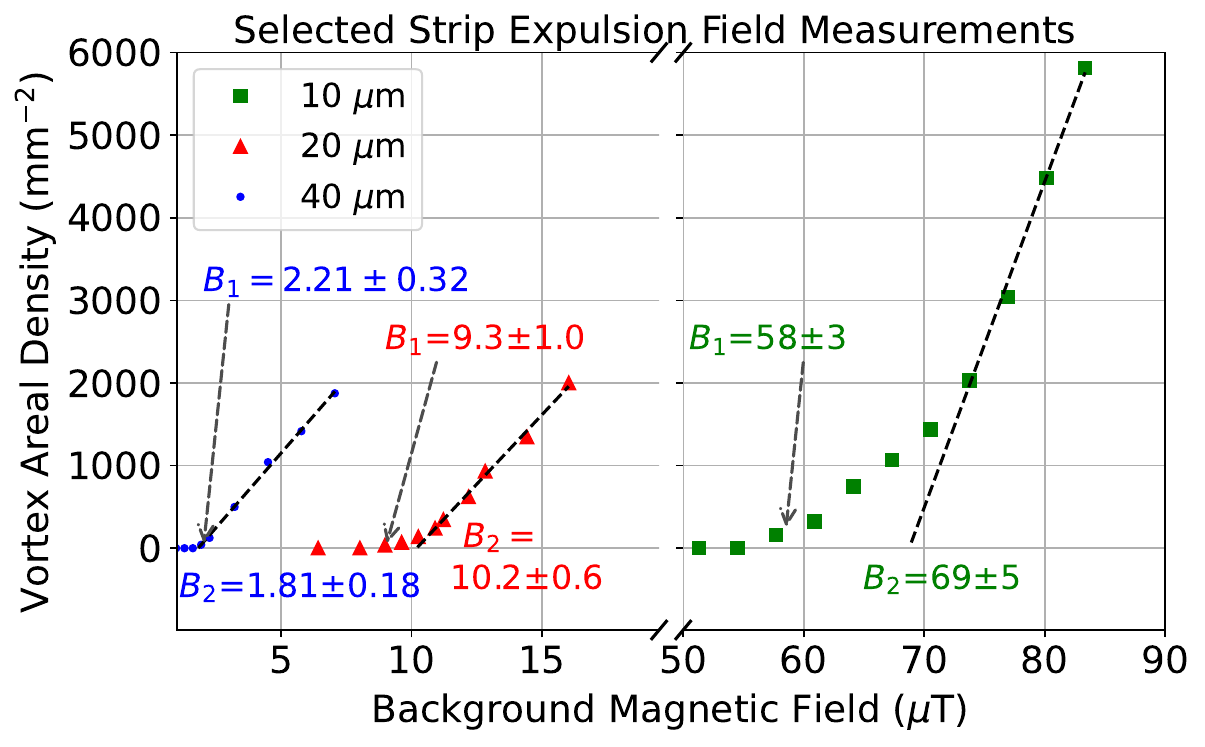}
        \put(2,62){\footnotesize\textbf{(a)}}
    \end{overpic}
    
        \vspace{-0.5em}
        \hspace*{0.1em}
    \begin{overpic}[width=.98\linewidth, trim=0pt 10pt 40pt 20pt, clip]{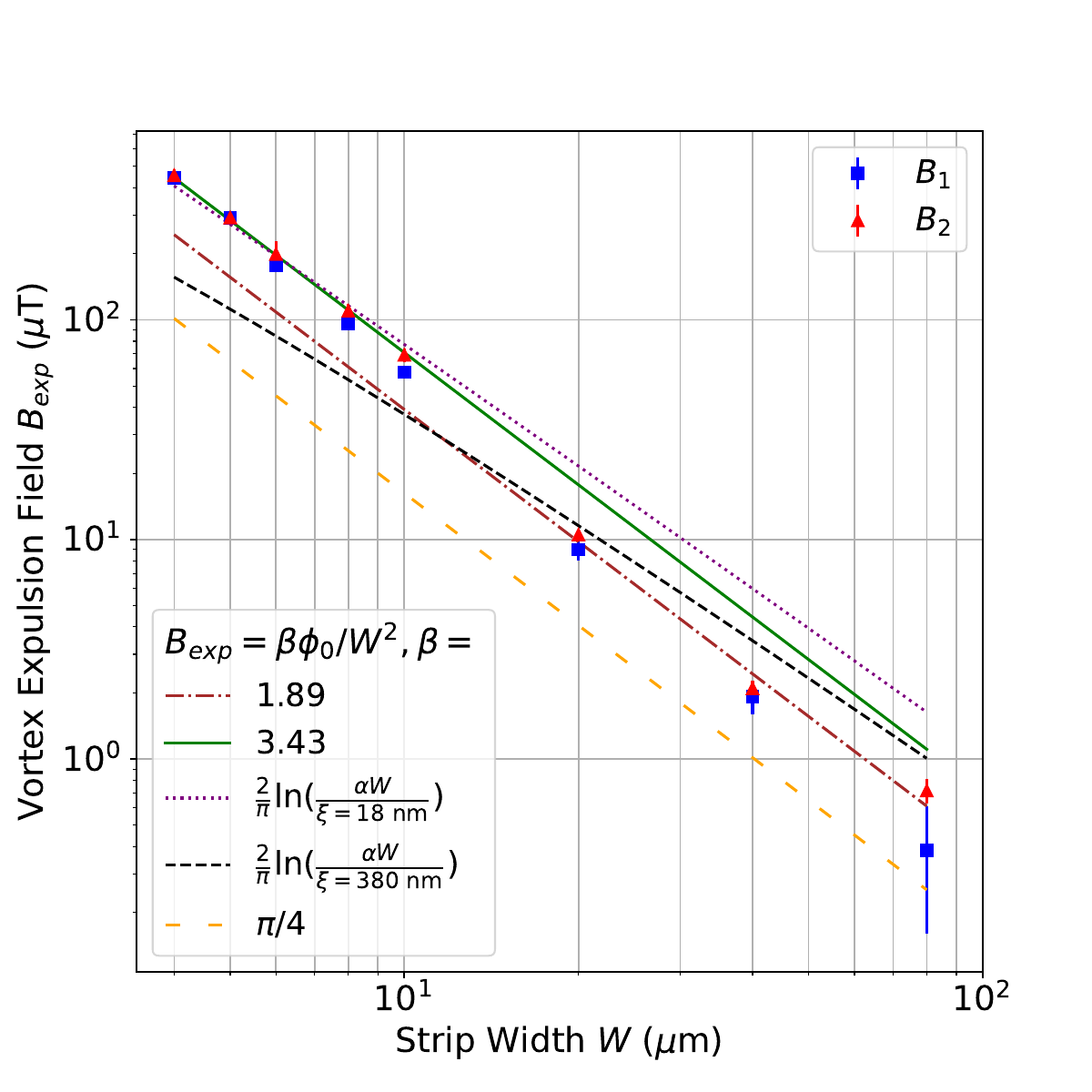}
        \put(2,92){\footnotesize\textbf{(b)}}
    \end{overpic}

    \caption{
(a) Vortex areal density $n_v$ vs.\ applied field $B_r$ for selected strip widths, showing the extracted expulsion fields $B_1$ and $B_2$ (in $\mu$T). Each $n_v$ value represents the number of vortices counted across multiple strips in a single cooldown (see Supplementary Table S2~\cite{suppl}). } (b) Expulsion field vs.\ strip width, expected to scale as $\beta \Phi_0/W^2$ with different predictions for the dimensionless factor $\beta$~\cite{washington1982observation}. Data for $W \leq 10~\upmu\text{m}$ and $W \geq 20~\upmu\text{m}$ agree well with $\beta = 3.43 \pm 0.12$ and $1.89 \pm 0.09$, respectively. The narrow-strip ($W \leq 10~\upmu\text{m}$) data match the theoretical $B_{c1}$ (Eq.~\ref{eq:Bc1}) for $\xi = 18 \pm 4$~nm, corresponding to Nb films at 4 K. The data deviate from $B_{c1}$ calculated using $\xi \approx 380$~nm near $T_c$. At all strip widths, the data exceed the vortex stability field $B_0 = \pi \Phi_0 / 4 W^2$ (Eq.~\ref{eq:eq1}).
    \label{fig:combined_critical_field_figures}
\end{figure}

The vortex expulsion field in a superconducting strip is expected to scale as $ B_{\mathrm{exp}} = \beta \Phi_0 / W^2 $, where $\beta$ is a geometry- and material-dependent numerical factor~\cite{washington1982observation}. Our measurements (Fig.~\ref{fig:combined_critical_field_figures}b) show that for $B_2$, $\beta = 3.43 \pm 0.12$ for narrow strips ($W \leq 10~\upmu\text{m}$), while $ \beta = 1.89 \pm 0.09$ for wider strips ($ W \geq 20~\upmu\text{m} $). We use $B_2$ rather than $B_1$ when fitting models to the expulsion field data, as $B_2$ corresponds to the expulsion field in an ideal film, while $B_1$ is more dependent on material quality, film defects, and non-idealities \cite{suppl}, though the $B_1$ and $B_2$ values in Fig.~\ref{fig:combined_critical_field_figures} are often similar.

Existing experimental results and theoretical models offer benchmarks for comparison of our results. For an ideal thin-film strip of infinite length, width $W$, thickness $d$, and magnetic field (London) penetration depth $\lambda$, the vortex energy in an applied field has been calculated in Ref.~\cite{likharev1971formation}, which found that:
\begin{enumerate}
    \renewcommand{\labelenumi}{\alph{enumi})}
    \item A vortex is energetically unfavorable at fields below
    \begin{equation}
        B_0 = \frac{\pi \Phi_0}{4W^2} \quad \left( \beta \leq \frac{\pi}{4}\right).  \label{eq:eq1}
    \end{equation}
    \item A vortex becomes energetically favorable at fields above
    \begin{equation}
        B_{c1} = \frac{2\Phi_0}{\pi W^2} \ln\left(\frac{\alpha W}{\xi(T)}\right) \quad (W \ll \Lambda(T)),
        \label{eq:Bc1}
    \end{equation}
    where \( \Lambda(T) = 2\lambda(T)^2/d \) is the temperature-dependent Pearl penetration depth, \( \xi(T) \) is the temperature-dependent coherence length, and $\alpha$ is an order-unity cutoff constant dependent on the assumed vortex size~\cite{initial_sc_strip_paper,likharev1971formation,fetterdiskBc}. Here we use $\alpha=2/\pi$~\cite{initial_sc_strip_paper}.
    \item A vortex is metastable for \( B_0 \leq B_r \leq B_{c1} \) in the potential well formed in the middle of the strip by its interaction with Meissner screening currents~\cite{likharev1971formation,bean1964surface}.
\end{enumerate}

Film defects can locally lower the vortex energy, creating vortex nucleation and pinning sites, and stabilizing vortices below $B_0$ or $B_{c1}$. Conversely, increased vortex energy\textemdash{}e.g., from suppressed screening along the film edges near $T_c$\textemdash{}can destabilize vortices at fields exceeding $B_0$. For example, Kuit \textit{et al.}~\cite{ybco_strip_critical_field, kuit2009vortex} observed no vortices in strips cooled in fields up to ${\sim}2B_0$, attributing this to interaction with thermally excited vortex-antivortex pairs near $T_c$~\cite{kosterlitz1973ordering}. In that framework, the effective expulsion field becomes:
\begin{equation}
    B_K = 1.65\frac{\Phi_0}{W^2} \quad (\beta = 1.65).
\end{equation}

Our measured $B_2$ values exceed $B_0$ and $B_K$ across all strip widths, and fall into two regimes. For $W > 20~\upmu\text{m}$, the data are consistent with $\beta=$ 1.89$\pm0.09$. For $W < 10~\upmu\text{m}$, the data fit yields an almost twice larger $\beta=3.43$, consistent with $B_{c1}$ (Eq.~\ref{eq:Bc1}) at a coherence length $\xi(T) = 18\pm4$~nm. This value of $\xi(T)$ is expected for Nb films at $T \approx 4$ K but is significantly smaller than the coherence length near $T_c$ ($\xi \approx 380~\text{nm}$ for $T_c - T \approx 0.02~\text{K}$), where vortex expulsion or freezing in the film is expected to occur. Figure~\ref{fig:combined_critical_field_figures}b includes a comparison to $B_{c1}$ calculated with both values of $\xi$. Disagreement between the large-$\xi$ $B_{c1}$ and the data may suggest that vortex core size at nucleation may be governed by the film's structural defects\textemdash{}e.g., grain boundaries or pores, which are predicted to be $\sim$20 nm and are a source of flux pinning in Nb films \cite{initial_sc_strip_paper, zerweck1981pinning, dasgupta1978flux, dew1987role, yetter1982grain, prokhorov1984pinning, PhysRevB.62.671, wu2015special} . 

The change in $B_2$ vs $W$ behavior between $W = 10~\upmu\text{m}$ and $W = 20~\upmu\text{m}$ may reflect film non-uniformities. Theoretical models assume homogeneous strips with uniform $T_c$ and superconducting parameters $\lambda$ and $\xi$ that vary uniformly with temperature. Under these assumptions, the criterion $W \ll \Lambda(T)$ can be satisfied across the entirety of the strip with any $W$ at temperatures arbitrarily close to $T_c$, as $\Lambda(T)$ diverges as:
\begin{equation}
    \Lambda(T) = \frac{2\lambda_0^2}{d} \left(1 - \frac{T^4}{T_c^4} \right)^{-1},
\end{equation}
where \( \lambda_0 = 85~\text{nm} \) and \( d = 200~\text{nm} \) for the Nb films in this work. However, real films exhibit spatial variation in $T_c$, with the resistive transition width of our films being $\Delta T_c \approx 30~\text{mK}$~\cite{suppl}. 

For narrow strips ($W\lesssim 10~\upmu$m), the condition $W\lesssim\Lambda(T)$ is satisfied at temperatures below $T_c - \Delta T_c/2$, after a coherent superconducting state has formed. In this regime, conventional vortex physics (Abrikosov or Pearl) is valid, as assumed in prior models~\cite{likharev1971formation, fetterdiskBc, koganringsBc, bean1964surface, maksimova1998mixed, initial_sc_strip_paper, kuit2009vortex, ybco_strip_critical_field}. 
In contrast, for wider strips ($W \gtrsim 20~\upmu\text{m}$), the condition $W\lesssim\Lambda(T)$ is met only during the resistive transition, before a uniform superconducting state is established. In this regime, the presence of vortex-antivortex pairs is possible, and flux may also be trapped via percolative mechanisms\textemdash{}e.g., through the formation of closed superconducting loops surrounding normal regions~\cite{washington1982observation}. This may explain the observed crossover in expulsion field scaling between 10 and 20~$\upmu$m widths, though further experiments are needed to fully understand this width-dependent behavior.

A difference between our measurements of expulsion field as a function of strip width and prior measurements and models is that, instead of considering a single strip, we used a set of parallel strips with spacing $s$ (3 to 20 $\upmu$m) for increased film area and better statistics (Fig.~\ref{fig:critical_field_measurement}). Theoretically, the presence of adjacent superconducting strips may affect the value of $B_{\mathrm{exp}}$ by providing an additional screening and enhancing magnetic field between the strips with respect to the applied field~\cite{BrandtIndenbomStrip, MawatariStrip}. Qualitatively, this effect should reduce the apparent $B_{\mathrm{exp}}$ with respect to a single strip, an effect that should depend on $s$ and should be more pronounced at large $W$. These dependences can be easily investigated using our imaging technique, and we leave this subject as the topic of a future study.

\section{Conclusions and Outlook}
We have demonstrated a widefield cryogenic NV-diamond magnetic microscope capable of rapid high-resolution imaging of magnetic flux (vortices) in superconducting films and devices. Its fast measurement rate enables the collection of statistically significant vortex formation data across diverse film geometries and cooldown cycles. Our measurements reveal that vortex expulsion fields in thin-film Nb strips scale as $\beta \Phi_0 / W^2$ with $\beta = 1.89$ for wider strips ($W \geq 20~\upmu\text{m}$), increasing to $\beta = 3.43$ for narrower strips ($W \leq 10~\upmu\text{m}$), corresponding to expulsion fields roughly twice those reported in prior studies~\cite{ybco_strip_critical_field, kirtley2010review}. This widefield magnetic imaging instrument opens new possibilities for investigating vortex dynamics and flux trapping in SCE, with direct relevance for scalable computing architectures.

Using MFI to solve the magnetic flux trapping challenge for SCE devices requires a reliable method that can measure large-area devices in a short time, allowing SCE technology developers to collect vortex formation statistics across multiple cooldown cycles.  As such, a key figure of merit for choosing an optimal MFI approach is the area measurement rate $\dot{A}$, which reflects the advantages of MFI instruments that can rapidly measure vortices with good SNR over a large area \cite{suppl}.  Our approach joins a variety of existing MFI techniques for imaging SCE devices, including scanning SQUID microscopy (SSM) \cite{kirtley2010review}, magneto-optic imaging (MOI) \cite{moiDOEref}, and others.   SSM instruments can exhibit exceptional single-pixel magnetic noise floors, but the area measurement rate is limited by the sensor scan speed, leading to typical area measurement rates of $\dot{A} \approx 23~\mathrm{to}~100~\upmu$m$^2$/s~\cite{suppl, vortexNbMOIprb2025}.  MOI has several things in common with NV magnetic imaging:  both use optical microscopy techniques to probe a magnetically sensitive slab, have a magnetic sensitivity set by the photon shot noise limit, and have a spatial resolution limited by the optical diffraction limit.  However, the MOI performance is about $\dot{A} \approx 30~\upmu$m$^2$/s~\cite{suppl, formFactorIQ1000}. With a combination of spatial resolution, sensitivity, and FOV size, the NV cryo-microscope compares favorably to other MFI approaches for studying magnetic flux trapping, with $\dot{A} \approx 860~\upmu$m$^2$/s in this work and room to improve.

Further measurements could include detailed characterization of films with various moat shapes and dimensions (Fig.~\ref{fig:critical_field_measurement}a), as well as broader investigations into chip architecture, such as the effects of moat geometry and spacing, and the influence of multiple superconducting layers \cite{nvMoatsPaper}. In parallel, upgrades to the diamond cryo-microscope setup are anticipated to further improve measurement throughput. Increasing the SNR and the FOV size would accelerate data collection over larger chip areas. Faster acquisition could also enable time-resolved imaging of vortex dynamics near $T_c$ or vortex movement resulting from flux shuttling \cite{ratchetEffect}. Additional enhancements could support imaging of active SCE circuits, allowing real-time flux imaging to be correlated with circuit performance or in-situ electrical diagnostics. Finally, implementing a pulsed version of the apparatus could extend its capabilities to the detection of MHz- and GHz-frequency currents in superconducting devices, offering broader diagnostic reach~\cite{samQFMimaging}.

\section{Acknowledgements}
We thank Tzu-Ming Lu and Andrew Mounce at the Sandia National Laboratories Center for Integrated Nanotechnologies for their assistance in preparing magnetic test samples for instrument performance validation and for providing diamond samples during instrument development. We thank Tom Osadchy for assistance with diamond growth; Peter O'Brien, Jon Wilson, and Matthew Ricci for help with diamond coating; and Anil Mankame and Tom Grasso for assistance with instrument assembly and testing. We are grateful to Ravi Rastogi and David Kim for overseeing fabrication of the Nb films and test structures used in this work, and to Rabindra Das for help with packaging. We thank Michele Kelley, Joseph Belarge, Andrew Wagner and Logan Bishop-Van Horn for helpful discussions. Rohan Kapur thanks John Kim for continued assistance.   This research was conducted with funding from the Under Secretary of Defense for Research and Engineering under Air Force Contract No.~FA8702-15-D-0001. The opinions, interpretations, conclusions, and recommendations are those of the authors and are not necessarily endorsed by the United States Government.  This work was performed, in part, at the Center for Integrated Nanotechnologies, an Office of Science User Facility operated for the U.S.~Department of Energy (DOE) Office of Science.

\appendix       
\setcounter{section}{0}

\section{Diamond Specifications} \label{diamondSpecsAppendix}
We used two diamond samples for this work. The first diamond was grown at MIT-LL via microwave plasma chemical vapor deposition on an electronic-grade substrate. The electronic-grade seed was mechanically polished to 0.3 mm thickness with a miscut angle of 3.4$\degree$ from the [100] crystallographic plane, then subjected to a brief MW plasma containing ultra-pure $^{12}$C methane ($>$99.999\%), $^{15}$N nitrogen (99.9\%), and H$_{2}$ ($>$99.999\%), resulting in $\sim$2~$\upmu$m of epitaxially grown nitrogen-doped diamond. The as-grown sample was polished to reduce the surface roughness and the NV layer thickness to $\sim$1 $\upmu$m, and was then electron irradiated, followed by annealing at 800 $\degree$C for 24 hours under vacuum to create the NV layer. The second diamond is a commercial research-grade diamond. It has a 1 $\upmu$m NV layer on an undoped 0.5 mm electronic-grade substrate cut along the [100] crystallographic plane. 

\section{Niobium Film Specifications}
Niobium films with 200 nm thickness were grown at room temperature by DC magnetron sputtering on 200 mm Si wafers covered by a 100 nm amorphous SiO$_2$ layer grown by a high-density plasma-enhanced chemical vapor deposition (PECVD). At these conditions, the Nb films form highly textured (110)-oriented columnar grains having a rice-grain shape in the film plane, with an average grain size of $\sim$50 nm. The films have $T_c = 9.2~\mathrm{K}$, determined as the midpoint of the resistive transition, with measured transition width $\Delta T_c = 30~\mathrm{mK}$. The sheet resistance at 295 K is $R_{295} = 0.95~\Omega$/sq, corresponding to a resistivity of $\rho = 18.0~\upmu \Omega \cdot \mathrm{cm}$. The residual-resistance ratio (RRR) is $R_{295}$/$R_{n} = 5.7$, where  $R_{n}$ is the sheet resistance just above $T_c$, taken at 9.3 K in this work. The residual resistivity of the films is about $\rho_0 = 3.2 ~\upmu \Omega \cdot \mathrm{cm}$ and the phonon resistivity at 295 K is $\rho_{ph}(295~\mathrm{K}) = \rho(295~\mathrm{K})- \rho_0 = 14.8~\upmu \Omega \cdot \mathrm{cm}$, which is consistent with the known phonon resistivity of bulk clean Nb $\rho_{Nb}(295~\mathrm{K}) = 14.7~\upmu \Omega \cdot \mathrm{cm}$.



%

\clearpage


\section*{Supplementary material}

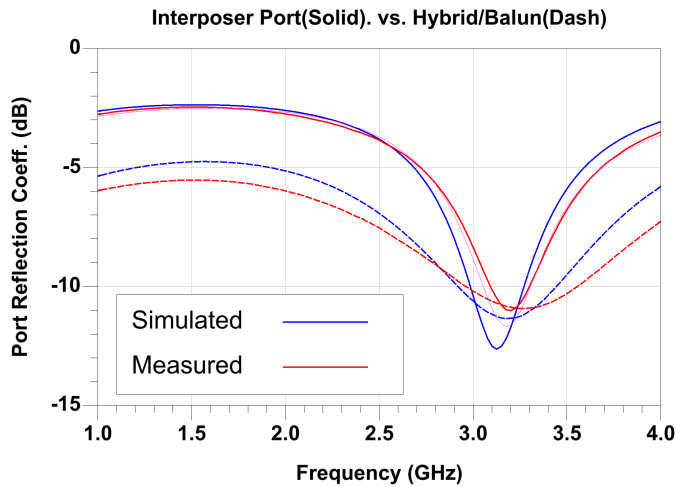

**Supplementary Figure** S1. Reflection coefficient (solid lines) of an interposer port with the other port terminated. Dashed lines show reflection coefficient of ports driven in antiphase through a 180° hybrid, which better reflects how the MINT is driven in the setup.

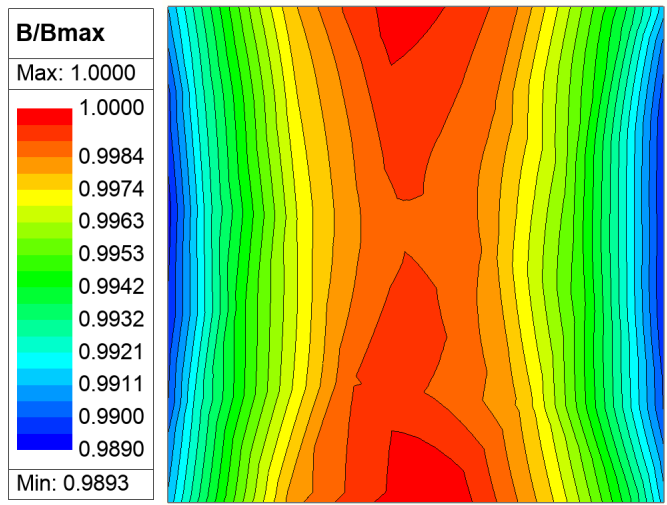

**Supplementary Figure** S2. B-field intensity within a $1\,\text{mm}^2$ field of view at the NV layer near the center of the diamond, normalized to the maximum.

interference (EMI) and dispersion, while an impedance transformer was implemented to reduce reflections from the diamond-interposer interface. Electrical continuity is maintained through mechanical contact between the MINT and the conductive coatings on the diamond. The diamond surface facing the optical window (the non-NV side) is coated with indium-tin-oxide (ITO) and $\text{Ta}_2\text{O}_5/\text{SiO}_2$, while the surface adjacent to the superconducting sample (the NV side) is coated with $\text{Al}/\text{Au}/\text{SiO}_2$. The microwave connection is made through the ITO layer, and the ground return occurs through the Au layer. This configuration sets up a quasi-TEM mode within the diamond that is largely confined between the ITO and Au layer. The feed design and resistivity of the ITO layer distribute AC current evenly across the surface, producing a lateral magnetic field in the NV layer that is both uniform and perpendicular to the direction of propagation.

Full-wave electromagnetic simulations carried out using ANSYS HFSS predict a strong and uniform MW magnetic field within the diamond NV layer (Fig. S1-S2). For a 1 mW input power, the expected field strength in the NV layer with ITO sheet resistance of $\sim 57\,\Omega/\text{sq}$ is $B_{MW} \approx 1.6\,\mu\text{T}$ over a $1 \times 1\,\text{mm}^2$ field of view with $\sim 1\%$ non-uniformity. Subsequent measurements of the fully assembled diamond-interposer device are in good agreement with simulations. Resistive loss in the ITO layer is the primary MW loss mechanism; however, measured heating of the sensor head $\approx 0.25\,\text{K}$ is well within the cooling power limit of the cryostat.

## IV. NV HAMILTONIAN NEAR ZERO MAGNETIC FIELD

The NV Hamiltonian (in units of frequency) can be written as

$$H = DS_z^2 + E(S_y^2 - S_x^2) + A_\parallel S_z I_z + \frac{A_\perp}{2}(S_+ I_- + S_- I_+) + \gamma \vec{B} \cdot \vec{S}, \tag{S1}$$

where $\vec{S}$ and $\vec{I}$ are the electron spin-1 and $^{15}$N nuclear spin-1/2 operators, $D$ and $E$ are zero-field splitting parameters, $A_\parallel$ and $A_\perp$ are longitudinal and transverse $^{15}$N hyperfine coupling parameters, $\gamma$ is the NV electron gyromagnetic ratio, and $\vec{B}$ is the magnetic field in the NV coordinate system. We can neglect the transverse hyperfine and the transverse magnetic field terms, leaving us with

$$H = DS_z^2 + E(S_y^2 - S_x^2) + A_\parallel S_z I_z + \gamma B_z S_z. \tag{S2}$$

Solving for the eigenvalues yields transition frequencies of $D \pm \sqrt{(A_\parallel/2 \pm \gamma B_z)^2 + E^2}$, which are plotted in Fig. 2b of the main text.

For small $B_z$, the $m_I = \pm 1/2$ hyperfine sublevels split into $m_S = \pm 1$ doublets with frequency shifts that are linear and symmetric in $B_z$, which manifests as an ODMR lineshape broadening. For large $B_z$ (i.e. $\gamma B_z \gg A_\parallel/2$, the conventional operating conditions for NV magnetometry with a bias magnetic field), the electron-spin Zeeman effect is the dominant term, and the $m_S = \pm 1$ sublevels are split into doublets by the $A_\parallel$ hyperfine term. In this regime, the frequency differences between $m_S = 0 \leftrightarrow \pm 1$ sublevels are used to determine the magnetic field.

Note that the diamond NV layers are oriented approximately normal to the [100] crystallographic direction, meaning that the N-V axes are ~35° out of plane. An out-of-plane magnetic field $B_Z$ in the lab frame corresponds to $B_z = B_Z/\sqrt{3}$ in the NV frame.

## V. VORTEX PINNING ACROSS MULTIPLE COOLDOWNS

As an extension to Fig. 2 in the main text, we include an animated GIF file showing flux trapping in an unpatterned Nb film across ten cooldowns for $B_r = 0.64$ µT.

Figure S3 shows that vortices consistently appear in the same locations in 10 µm Nb strips across multiple cooldowns for $B_r = B_1$. This suggests that vortices may be trapped on pinning sites for $B_1 < B_r < B_2$.

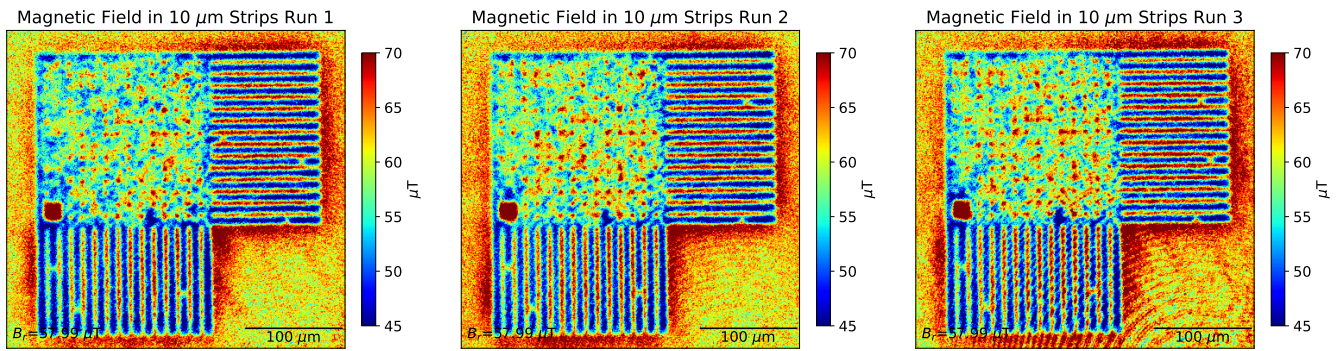

**Supplementary Figure** S3. Vortex pinning measurements across three cooldown cycles for Nb strips of width $W = 10$ µm and $B_r = 57.99$ µT, where $B_1 = 57.99$ µT and $B_2 = 69.22$ µT. The concentric rings in the bottom right corner are an artifact likely arising from dust on an optical element in the microscope.

 

\end{document}